\documentclass{PoS}
\usepackage{axodraw}
\usepackage{epsfig}

\def\beq{\begin{equation}}
\def\eeq{\end{equation}}
\def\bea{\begin{eqnarray}}
\def\eea{\end{eqnarray}}

\def\Eq#1{Eq.~(\ref{#1})}

\title{Tevatron anomalies and LHC cross-checks}

\ShortTitle{Tevatron anomalies and LHC cross-checks}

\author{\speaker{Germ\'an Rodrigo}\thanks{Supported by REA 
Grant Agreement PITN-GA-2010-264564 (LHCPhenoNet), 
by the MICINN Grant No. FPA2007-60323, and PR2010-0481, 
by CPAN (Grant No. CSD2007-00042), and
by the Generalitat Valenciana Grant No. PROMETEO/2008/069.
G.R. acknowledges hospitality at the Institut f\"ur Theoretische 
Teilchenphysik of the Karlsruher Institut f\"ur Technologie 
during the completion of this work, and J. H. K\"uhn for a 
fruitful collaboration.}\\
Instituto de F\'{\i}sica Corpuscular, \\
Consejo Superior de Investigaciones Cient\'ificas-Universitat de Val\`encia, \\
Parc Cient\'ific, E-46980 Paterna (Valencia), Spain.
\\
        E-mail: \email{german.rodrigo@csic.es}}

\abstract{A sizeable charge asymmetry in top quark pair 
production has been observed at the Tevatron. 
The experimental results seem to exceed systematically the 
Standard Model theory predictions by a significant amount and have 
triggered a large number of suggestions for 'new physics'. 
The effect is also visible at the LHC, and preliminary 
results have already been presented by the ATLAS and CMS collaborations. 
In this talk, we review the present status of the theoretical 
predictions, and their comparison with the experimental measurements.}

\FullConference{The 2011 Europhysics Conference on High Energy Physics-HEP 2011,\\
		July 21-27, 2011\\
		Grenoble, Rh\^one-Alpes, France}

\begin{document}

\section{Introduction}

Top quark production at hadron colliders is one of the most active
fields of current theoretical and experimental studies~\cite{Galtieri:2011yd},
and most probably the most promising probe of physics beyond 
the Standard Model (SM).
Sizable differences have been observed between theory 
predictions~\cite{Antunano:2007da,Kuhn:1998kw,Kuhn:1998jr,Bowen:2005ap} 
for the top quark charge asymmetry and measurements by the CDF and the 
D0 collaborations~\cite{Abazov:2011rq,d0,Aaltonen:2011kc,cdf19} 
at the Tevatron. 
The discrepancy is particularly pronounced for the subsample of
$t\bar t$ pairs with large invariant mass, $m_{t\bar t} > 450$~GeV,
and the asymmetry defined in the $t\bar t$ rest frame, 
where a $3.4 \sigma$ effect has been claimed~\cite{Aaltonen:2011kc}. 
It is interesting to note, however, that the discrepancy is less 
prominent in the laboratory frame~\cite{Aaltonen:2011kc}.
The D0 Collaboration also finds positive 
discrepancies with the SM~\cite{Abazov:2011rq}.
These discrepancies have triggered a large number of theoretical 
investigations, using these
results, either to restrict new physics like heavy 
axigluons~\cite{Ferrario:2009bz,Rodrigo:2010gm} or to
postulate a variety of new phenomena in 
the t-channel~\cite{Jung:2009jz,Cheung:2009ch,Shu:2009xf}.
At the same time, the robustness of the leading order QCD prediction 
has been studied in~\cite{Almeida:2008ug,Ahrens:2010zv},
where it has been argued that  next-to-leading (NLL) as well as
next-to-next-to leading (NNLL) logarithmic corrections do not
significantly modify the leading order result, in agreement with the
approach advocated in~\cite{Kuhn:1998kw,Kuhn:1998jr} 
(Note, however, the large corrections observed in Ref.~\cite{Dittmaier:2008uj} 
for the corresponding studies of the $t\bar t$+jet sample). 
A small modification of the SM prediction arises from inclusion of QED
corrections. In Ref.~\cite{Kuhn:1998kw} this effect was estimated 
to lead to an increase of the the QCD asymmetry by a factor 1.09, in recent 
analysis~\cite{arXiv:1109.6830,Hollik:2011ps}, 
however, an enhancement factor of 1.2 has been obtained.
Obviously this small increase of the SM prediction for the asymmetry
cannot resolve the discrepancy between theory an experiment mentioned
above. 

In this talk we revisit the SM prediction of the top quark charge 
asymmetry at the Tevatron and the LHC~\cite{arXiv:1109.6830}.  
We paid special attention to the electroweak corrections.
We summarize the experimental measurements of the asymmetry and 
update the pull of their discrepancy with the SM. 
We also analyze the effect of introducing a cut in the 
$t\bar t$ transverse momentum (see also~\cite{Demina}). 
Finally, we introduce a new 
quantity $A_{t\bar t}(Y)$, which measures the 
forward--backward asymmetry with respect to the average rapidity 
of top and antitop quarks, being a suitable observable 
both at the Tevatron and the LHC.
Beyond the SM contributions to the asymmetry have also been discussed
in this conference in~\cite{Westhoff}, and will not be covered in 
this document. 

\section{The charge asymmetry in the SM}

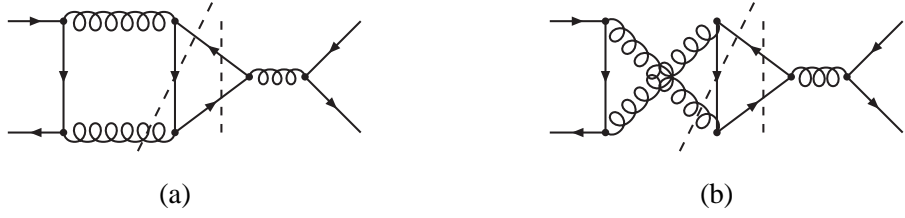
\begin{figure}[htb]
\begin{center}

\begin{picture}(200,70)(0,0)
\SetWidth{1.1}
\SetScale{.7}

\SetOffset(-10,40)

\Vertex(40,0){2}
\Vertex(70,0){2}
\Vertex(0,30){2}
\Vertex(0,-30){2}
\Vertex(-60,30){2}
\Vertex(-60,-30){2}
\Gluon(-60,30)(0,30){5}{6}
\Gluon(0,-30)(-60,-30){5}{6}
\Gluon(40,0)(70,0){4}{3}
\ArrowLine(0,30)(0,-30)
\ArrowLine(0,-30)(40,0)
\ArrowLine(40,0)(0,30)
\ArrowLine(100,30)(70,0)
\ArrowLine(70,0)(100,-30)
\ArrowLine(-90,30)(-60,30)
\ArrowLine(-60,30)(-60,-30)
\ArrowLine(-60,-30)(-90,-30)
\DashLine(25,30)(25,-30){5}
\DashLine(20,40)(-20,-40){5}
\Text(0,-45)[]{(a)}

\SetOffset(195,40)

\Vertex(40,0){2}
\Vertex(70,0){2}
\Vertex(0,30){2}
\Vertex(0,-30){2}
\Vertex(-60,30){2}
\Vertex(-60,-30){2}
\Gluon(-60,30)(0,-30){5}{8}
\Gluon(0,30)(-60,-30){5}{8}
\Gluon(40,0)(70,0){5}{3}
\ArrowLine(0,30)(0,-30)
\ArrowLine(0,-30)(40,0)
\ArrowLine(40,0)(0,30)
\ArrowLine(100,30)(70,0)
\ArrowLine(70,0)(100,-30)
\ArrowLine(-90,30)(-60,30)
\ArrowLine(-60,30)(-60,-30)
\ArrowLine(-60,-30)(-90,-30)
\DashLine(25,30)(25,-30){5}
\DashLine(20,40)(-20,-40){5}
\Text(0,-45)[]{(b)}

\end{picture}
\end{center}
\caption{Cut diagrams representing the QCD contribution to the charge asymmetry.\label{fig:cut}}
\end{figure}

The dominant contribution to the charge asymmetry
originates from $q\bar{q}$ annihilation~\cite{Kuhn:1998kw,Kuhn:1998jr}. 
Specifically, it originates from the interference between the 
Born amplitudes for $q\bar{q}\to Q\bar{Q}$ and the part of the one-loop 
correction, which is antisymmetric under the exchange of quark 
and antiquark (box and crossed box). 
To compensate the infrared divergences, this virtual correction 
must be combined with the interference between initial and 
final state radiation.
Diagrams with triple gluon coupling in both real and virtual 
corrections give rise to symmetric amplitudes
and can be ignored. The corresponding contribution to the rate is conveniently 
expressed by the absorptive contributions (cuts) of the 
diagrams depicted in Fig~\ref{fig:cut}.
A second contribution to the asymmetry from 
quark-gluon scattering (``flavor excitation'') hardly contributes to 
the asymmetry at the Tevatron. At the LHC, it enhances 
the asymmetry in suitable chosen kinematical regions~\cite{Kuhn:1998jr}.

%%%%%%%%%%%%%%%%%%%%%%%%%%%%%%%%%%%%%%%%%%%%%%%
%%%%%%%%%%%%%%%%%%%%%%%%%%%%%%%%%%%%%%%%%%%%%%%
%%%%%%%%%%%%%%%%%%%%%%%%%%%%%%%%%%%%%%%%%%%%%%%
%%%%%%%%%%%%%%%%%%%%%%%%%%%%%%%%%%%%%%%%%%%%%%%

Diagrams similar to those depicted in Fig.~\ref{fig:cut}, where
one of the gluons has been substituted by a photon, also lead 
to a contribution to the charge asymmetry from mixed QED-QCD
corrections. The relative factor between QCD and QED 
asymmetries amounts to 
\beq
f_q^{\rm QED} = 3 \, \frac{\alpha_{\rm QED} \, Q_t \, Q_q}
{\displaystyle \frac{\alpha_S}{2} \, \left( \frac{d_{abc}^2}{4}\right)^2}
= \frac{\alpha_{\rm QED}}{\alpha_S} \, \frac{36}{5} \, Q_t \, Q_q
\label{eq:fqQED}
\eeq
for one quark species, and to 
\beq
f^{\rm QED} = \frac{4 f_u^{\rm QED} +  f_d^{\rm QED}}{5} = 
\frac{\alpha_{\rm QED}}{\alpha_S} \, \frac{56}{25} \approx 0.18~,
\label{eq:QED}
\eeq
after convolution with the PDFs if one considers as a first approximation 
that the relative importance of $u\bar u$ versus $d\bar d$ 
annihilation at the Tevatron is $4:1$. 
Thus, to an enhancement of nearly twenty percent 
of the QCD asymmetry, in good agreement with the more detailed 
numerical studies of~\cite{arXiv:1109.6830,Hollik:2011ps}.
At the LHC, the relative importance of $u\bar u$ versus $d\bar d$ 
annihilation is approximately $2:1$, thus reducing $f^{\rm QED}$ by a 
factor $5/7$ down to $0.13$. Similarly, weak contributions with the 
photon replaced by the $Z$ boson should be considered at the same 
footing. However, as a consequence of the cancellation between up and 
down quark contributions, and the smallness of the weak coupling, 
the weak corrections at the Tevatron are smaller by more than a factor 
$10$ than the corresponding QED result. 
For proton-proton collisions the cancellation between up and down 
quark contributions is even stronger and the total weak correction 
is completely negligible.

\section{Tevatron}

\begin{figure}[t]
\begin{center}
\includegraphics[width=6cm]{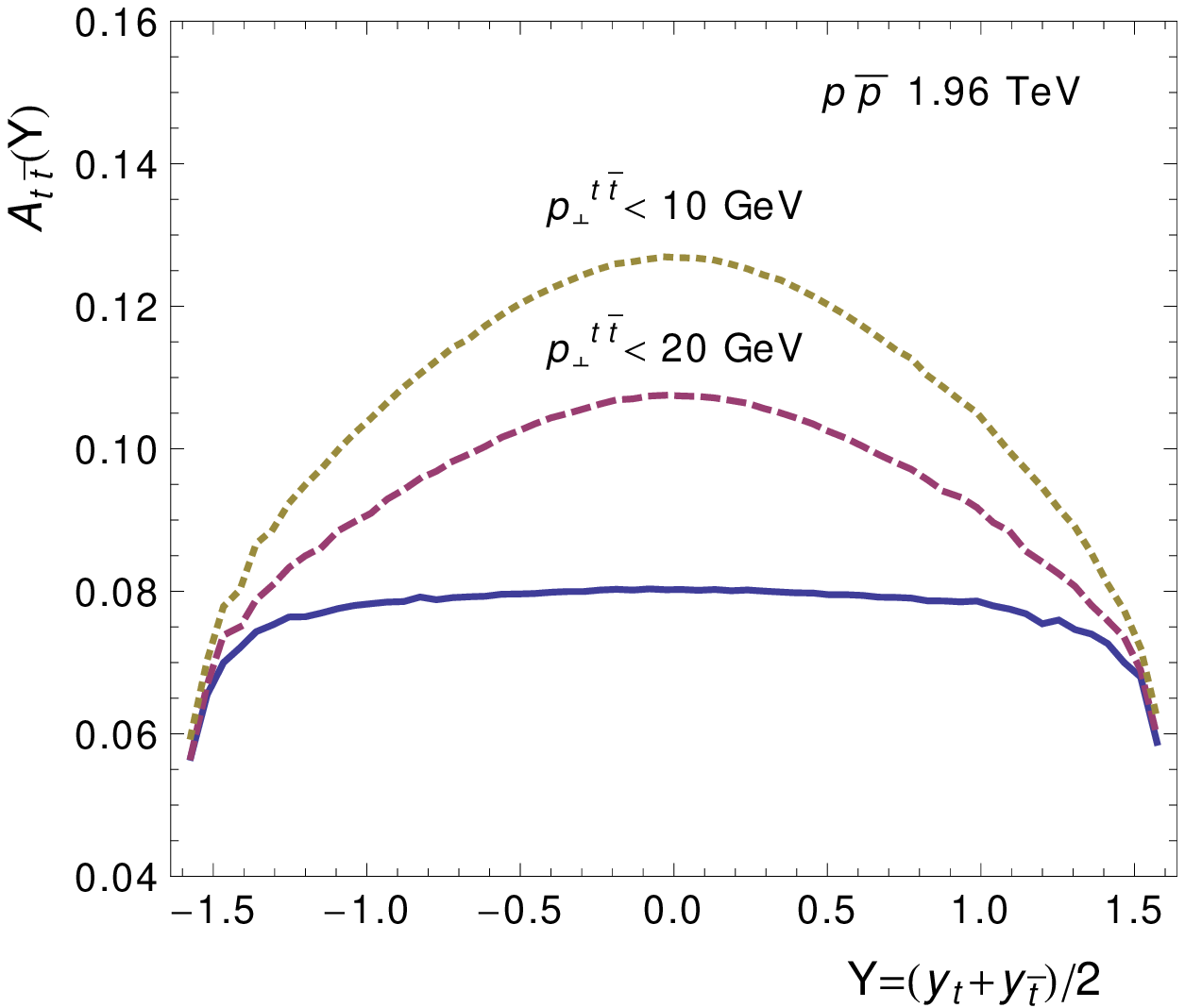} \qquad
\includegraphics[width=6cm]{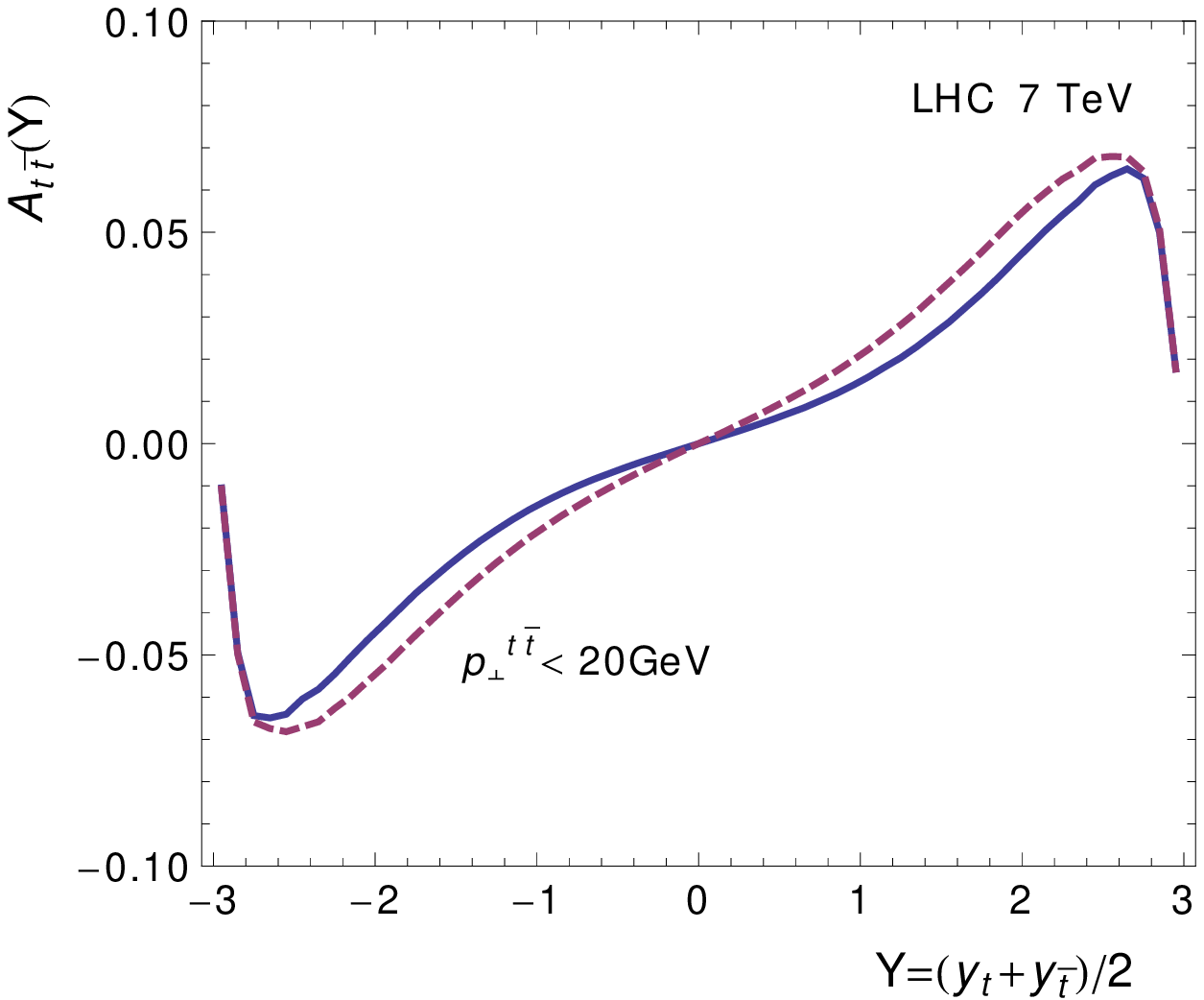}
\caption{Pair charge asymmetry $A_{t\bar t} (Y)$ as a function of  
the mean rapidity $Y=(y_t+y_{\bar t})/2$. 
Solid line: without cut on $p_\perp^{t\bar t}$, 
dotted/dashed lines: with cut on $p_\perp^{t\bar t}$.
\label{fig:AY}}
\end{center}
\end{figure}

Assuming that the rapidities of $t$ and $\bar t$ have 
been measured simultaneously, one defines the asymmetry
\beq
A_{t\bar t}\, (Y)=\frac{N(y_t>y_{\bar t})-N(y_{\bar t}>y_t)}
{N(y_t>y_{\bar t})+N(y_{\bar t}>y_t)}~,
\label{eq:pair}
\eeq
where $Y=(y_t+y_{\bar t})/2$ has been fixed. 
An almost flat asymmetry $A_{t\bar t} (Y)$ of around $8\%$ is predicted
at Tevatron as a function of $Y$ (Fig.~\ref{fig:AY} left). 
Two versions of the integrated asymmetry have been introduced
in Refs.~\cite{Antunano:2007da,Kuhn:1998kw,Kuhn:1998jr}: 
the forward--backward asymmetry in the laboratory frame
\beq
A_{\rm lab}=\frac{N(y_t>0)-N(y_t<0)}{N(y_t>0)+N(y_t<0)}
= \frac{N(y_t>0)-N(y_{\bar t}>0)}{N(y_t>0)+N(y_{\bar t}>0)}~,
\eeq
and the asymmetry in the $t\bar t$ rest frame
\beq
A_{t\bar t}=\frac{N(y_t>y_{\bar t})-N(y_{\bar t}>y_t)}
{N(y_t>y_{\bar t})+N(y_{\bar t}>y_t)}~.
\eeq
Results for both of them in the SM are listed in Table~\ref{tab:Attbar}. 
These predictions include also the QED and weak (strongly suppressed) 
corrections. Those corrections enhance the QCD asymmetry by  
an overall factor $1.21$, which is slightly different 
from \Eq{eq:QED} due to the deviation of the relative amount of 
$u\bar u$ and $d\bar d$ contributions from the simple approximation $4:1$. 

%%%%%%%%%%%%%%%
%%%%%%%%%%%%%%%

%%%%%%%%%%%%%%%
\begin{table}[t]
\begin{center}
\caption{Predicted asymmetries in the laboratory $A_{\rm lab}$
and the $t\bar t$ rest frame $A_{t \bar t}$ at Tevatron. 
Predictions are given also for 
samples with the top quark pair invariant mass $m_{t\bar t}$
above and below $450$~GeV, and with $|\Delta y|=|y_t-y_{\bar t}|$ larger 
or smaller than one in the $t\bar t$ rest frame. \label{tab:Attbar}}

\begin{tabular}{|c|ccccc|} \hline
laboratory     
          & $A_{\rm lab}$ & $m_{t\bar t}< 450$ GeV & $m_{t\bar t}> 450$ GeV & &  \\ \hline 
SM             & 0.056 (7) & 0.029 (2) & 0.102 (9) & & \\ \hline
MCFM~\cite{Aaltonen:2011kc}         
               & 0.038 (6)   &  &  & & \\ \hline\hline
$t\bar t$ rest frame
          & $A_{t \bar t}$ & $m_{t\bar t}< 450$ GeV & $m_{t\bar t}> 450$ GeV 
          & $|\Delta y|<1$ & $|\Delta y|>1$\\ \hline 
SM             & 0.087 (10)& 0.062 (4) & 0.128 (11) 
               & 0.057 (4) & 0.193 (15) \\ \hline
MCFM~\cite{Aaltonen:2011kc}         
               & 0.058 (9) & 0.040 (6)   & 0.088 (13) 
               & 0.039 (6) & 0.123 (18)\\ \hline
\end{tabular}
\end{center}
\end{table}
%%%%%%%%%%%%%%%

In order to compare theoretical results in the SM with 
the most recent measurements at Tevatron,
predictions in Table~\ref{tab:Attbar} are presented also for samples
with $m_{t\bar t}$ larger and smaller than $450$~GeV, and 
with $|\Delta y| = |y_t-y_{\bar t}|$ larger and smaller than $1$.
It is also interesting to compare these results with those based 
on a Monte Carlo prediction~\cite{Aaltonen:2011kc} 
based on MCFM~\cite{Campbell:1999ah}.
The enhancement factor of the SM result in Table~\ref{tab:Attbar}
compared to MCFM of about $1.5$ is easily understood: a 
factor $1.2$ originates from the inclusion of QED effects. 
Another factor of about $1.3$ originates from normalizing with 
respect to the Born cross-section instead of the NLO result. 
Since the asymmetric part of the cross-section is presently 
known to LO only we consider the normalization to the LO
cross-section more 
plausible~\cite{Kuhn:1998kw,Kuhn:1998jr,Almeida:2008ug,Ahrens:2010zv}.

%%%%%%%%%%%%%%%
\begin{figure}[ht]
\begin{center}
\includegraphics[width=6cm]{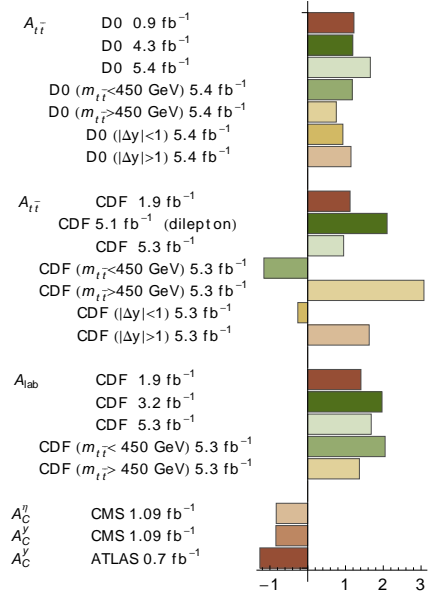}
\caption{Summary of experimental measurements of the charge asymmetry 
in comparison with the SM theoretical predictions. The histogram 
represents the pull of the discrepancy
for each measurement.
\label{fig:thexp}}
\end{center}
\end{figure}
%%%%%%%%%%%%%%%

A graphical illustration of the results in terms 
of the ''pull'' (measured in standard deviations) is shown in 
Fig.~\ref{fig:thexp}. 
The systematic upward shift of all but two Tevatron results is evident. 
The highest discrepancy, as has extensively been discussed in the 
literature, occurs for samples with $m_{t\bar t} > 450$~GeV and the charge 
asymmetry defined in the $t\bar t$ rest frame. 
Also shown in this Figure are preliminary results 
from CMS~\cite{CMS} and ATLAS~\cite{ATLAS} 
with a slight pull in the opposite direction.

%%%%%%%%%%%%%%%%%%%%%%%%%%%%%%%%%%%%%%%%%%%%%%%%%%%%%%%%%%%%%

The impact of cuts on hard gluon (and photon)
radiation on $A_{t\bar t} (Y)$ is also shown in Fig.~\ref{fig:AY}.
The dotted and dashed curves in Fig.~\ref{fig:AY}
show the effect of a cut on $p_\perp^{t\bar t}$ for values of 
$p_\perp^{\rm max} = 10$~GeV and $20$~GeV, respectively. 
An increase of the asymmetry by more than a factor 1.5 in the central 
region is observed for the most restrictive choice of $10$~GeV, 
and even a fairly loose $p_\perp^{\rm max} = 20$~GeV
modifies the asymmetry by up to a factor 1.3. 

\section{LHC}

The charge asymmetry can also be investigated in proton-proton
collisions at the LHC~\cite{Antunano:2007da,Kuhn:1998kw,Kuhn:1998jr} 
by exploiting the small $t\bar t$ sample produced in annihilation 
of valence quarks and antiquarks from the sea. Since valence quarks carry 
on average more momentum than sea antiquarks,
production of top quarks with larger rapidities will be preferred in the SM, 
and antitop quarks will be produced more frequently at smaller 
rapidities. This observation suggests to define 
the cut-dependent asymmetries 
\beq
A_C^{\rm in}(y_C) = \frac{N(|y_{\bar t}|\le y_C)-N(|y_t|\le y_C)}
{N(|y_t|\le y_C)+N(|y_{\bar t}|\le y_C)}
\label{eq:aCin}
%\eeq
\qquad {\rm and} \qquad 
%\beq
A_C^{\rm out}(y_C) = \frac{N(|y_t|>y_C)-N(|y_{\bar t}|>y_C)}
{N(|y_t|>y_C)+N(|y_{\bar t}|>y_C)}~,
\eeq
which serve to characterize the depletion of top quarks in the 
central region ($A_C^{\rm in}(y_C)>A_C^{\rm out}(y_C)$ for $y_C \lesssim 0.7$ 
approximately~\cite{Antunano:2007da,Ferrario:2008wm,Ferrario:2009ee}), 
and their enhancement at larger rapidities; 
$A_C^{\rm out}$ is much larger than $A_C^{\rm in}$ at large values of 
the rapidity cut $y_C$~\cite{arXiv:1109.6830}. 
This is because the central region is dominated by gluon fusion 
processes, while the sample with large rapidities has a larger 
relative content of $q\bar q$ initiated events. 
The statistical significance of both observables is, however, 
very similar~\cite{Hewett:2011wz} because the larger size of the asymmetry 
$A_C^{\rm out}$ with respect to $A_C^{\rm in}$
is compensated by the lower rate of events at larger rapidities.

The recent CMS~\cite{CMS} and ALTAS~\cite{ATLAS} analysis
have considered also the cut-independent charge asymmetries 
\beq
A_C^\eta = \frac{N(\Delta_\eta>0)-N(\Delta_\eta<0)}
{N(\Delta_\eta>0)+N(\Delta_\eta<0)} \qquad {\rm and} \qquad
A_C^y = \frac{N(\Delta_y>0)-N(\Delta_y<0)}
{N(\Delta_y>0)+N(\Delta_y<0)}~,
\eeq
where $\Delta_\eta = |\eta_t|-|\eta_{\bar t}|$ and 
$\Delta_y = |y_t|-|y_{\bar t}|$ or $y_t^2-y_{\bar t}^2$.
The SM predictions for the integrated asymmetries are 
listed Table~\ref{tab:AttbarmultiTeV} for different center-of-mass 
energies of the LHC, together with the experimental results for 
$\sqrt{s}=7$~TeV. Both experiments obtain negative asymmetries, 
although compatible with the SM prediction within uncertainties. 
New analysis with larger statistics are underway. 

%%%%%%%%%%%%%%%%%%%%%%%%%%%%%%%%%%%%%%%%%%%%%%%%%%%%%%%%%%%%%

\begin{table}[t]
\begin{center}
\caption{SM cut-independent charge asymmetries $A_\eta$  and $A_y$,
and integrated pair charge asymmetry $A_{t\bar t}^{cut}(Y_{\rm cut}=0.7)$, 
at different LHC energies. Summary of recent measurements by CMS and ATLAS. 
\label{tab:AttbarmultiTeV}}

\begin{tabular}{|l|ccc|} \hline     
               & $A_C^\eta$   & $A_C^y$     &  $A_{t\bar t}^{\rm cut}(Y_{\rm cut}=0.7)$  \\ \hline 
LHC 7 TeV      & 0.0136 (8)  & 0.0115 (6) &  0.0203 (8)  \\
%LHC 8 TeV      & 0.0122 (7)  & 0.0102 (5) &  0.0178 (6)  \\
%LHC 10 TeV     & 0.0101 (6)  & 0.0082 (4) &  0.0142 (5)  \\
%LHC 12 TeV     & 0.0087 (5)  & 0.0068 (3) &  0.0117 (4)  \\
LHC 14 TeV     & 0.0077 (4)  & 0.0059 (3) &  0.0100 (4)  \\\hline\hline
LHC 7 TeV CMS~\cite{CMS} 
 & -0.016 $\pm$ 0.030 ${}^{+0.010}_{-0.019}$ 
 & -0.013 $\pm$ 0.026 ${}^{+0.026}_{-0.021}$  & \\
LHC 7 TeV ATLAS~\cite{ATLAS} &   
 & -0.024 $\pm$ 0.016 $\pm$ 0.023   & \\\hline
\end{tabular}
\end{center}
\end{table}

%%%%%%%%%%%%%%%%%%%%%%%%%%%%%%%%%%%%%%%%%%%%%%%%%%%%%%%%%%%%%

Top quark production in proton-proton collisions is dominated 
by gluon fusion, which, in turn, is dominant in the central region. 
Conversely, quark-antiquark annihilation will be more enriched
for events with $t\bar t$ at larger rapidities (and larger 
$m_{t\bar t}$). 
This suggest to employ the definition of~\Eq{eq:pair}, which 
is essentially the asymmetry in the $t\bar t$ rest frame, also 
for the present case, and concentrate on $t\bar t$ events at 
large rapidities. The prediction for $A_{t\bar t} (Y)$ is shown 
in Fig.~\ref{fig:AY} for $\sqrt{s}=7$~TeV (right plot).
By construction, $A_{t\bar t} (Y)$ is now 
an antisymmetric function of $Y$.
Since most of the charge asymmetry is concentrated at large rapidities 
the statistical significance of any measurement will be enhanced, 
if the sample is restricted to larger rapidities.
Let us therefore define the quantity
\beq
A_{t\bar t}^{\rm cut} \, (Y_{\rm cut})=\frac{N(y_t>y_{\bar t})-N(y_{\bar t}>y_t)}
{N(y_t>y_{\bar t})+N(y_{\bar t}>y_t)}~,
\label{eq:paircut}
\eeq
where $Y>Y_{\rm cut}$.
Theoretical predictions in the SM for $A_{t\bar t}^{\rm cut} \, (Y_{\rm cut}=0.7)$ 
are presented in Table~\ref{tab:AttbarmultiTeV}.
QED and weak corrections amount to roughly a factor 1.1.

\section{Summary}

Tevatron has shown in the last years a systematic upward discrepancy 
in the measurement of the top quark charge asymmetry with respect 
to theoretical predictions in the SM. These discrepancies have 
triggered a large number of theoretical speculations about 
possible contributions beyond the SM. The Tevatron collaborations 
can still increase the statistical significance of their measurements, 
particularly by combining CDF and D0 results. On the other hand, 
the LHC, due to his present good performance, will provide quite 
soon competitive and accurate measurements of this effect.

\end{document}